# Generating Digital Models Using Text-to-3D and Image-to-3D Prompts: Critical Case Study

Rushan Ziatdinov
*Department of Industrial Engineering, College of Engineering*
*Keimyung University*
704-701 Daegu, Republic of Korea
ziatdinov@kmu.ac.kr, ziatdinov.rushan@gmail.com

Rifkat Nabiyev
*Department of Ground Transport Operations in the Oil, Gas, and Construction Industries, Institute of Architecture and Civil Engineering*
*Ufa State Petroleum Technological University*
450064 Ufa, Russia
dizain55@yandex.ru

*Abstract*— In the world of technology and AI, digital models play an important role in our lives and are an essential part of the digital twins of real-world objects. They can be created by designers, artists, or game developers using spline curves and surfaces, meshes, and voxels, but making such models is too time-consuming. With the growth of AI tools, there is interest in the automated generation of 3D models, such as generative design approaches, which can save creators valuable time. This paper reviews several online 3D model generators and critically analyses the results, hoping to see higher-quality results from different prompts.

## I. Introduction

Digital twins [1] of real-world objects include 3D models created using geometric modelling or computer graphics tools such as splines [2], meshes, voxels, etc. They are an essential part of modern architecture, art, automotive design, augmented reality (AR), biology (e.g. medical models), cartography (e.g., 3D maps), cinema and animation, civil engineering, computer-aided design (CAD), cultural heritage, education and training, engineering and product design, entertainment and animation, fashion and jewellery design, film and visual effects, forensics and crime scene reconstruction, games, geology and geophysics, graphic design, healthcare and medical imaging, industrial design, interactive media, interior design, manufacturing and prototyping, marketing and advertising, product design, robotics, scientific visualisation, simulation and analysis, game development, virtual prototyping, and VR worlds and experiences.

Creating three-dimensional objects using curves, surfaces, and meshes requires the efforts of specialists, including artists, designers, shape modellers, colourists, quality control engineers, rendering specialists, and others. These specialists often focus on repetitive tasks that are used across many designs.

One of the most recent advances in research is using various artificial intelligence (AI) tools, such as various AI-based chatbots, image and video generators, website creators, and others. Less well-known tools include text-to-3D content creators, often known as 3D model generators. These include various online tools such as

- https://text-to-cad.zoo.dev/ (generates B-Rep CAD files and meshes from natural language prompts)
- https://nething.xyz/arena (image to 3D and generating 3D models through coding)
- https://www.gnucleus.ai/ (transforms text and images into CAD models by GenAI)
- https://www.meshy.ai/ (AI 3D model generator)

Other tools have been discussed in recent articles, such as Magic3D [3], DreamFusion [4], Fantasia3D [5], ProlificDreamer [6], Dream3D [7], Make-it-3D [8], One-2-3-45 [9], DreamBooth3D [10], Control3D [11], DreamHuman [12]. Still, they may not be easily accessible online and require installation and programming language use.

Creating 3D models from text and images is a relatively new topic. The first manuscript found by searching all fields ALL("Text-to-3D" OR "Text to 3D" OR "Image-to-3D" OR "Image to 3D" OR "3D Model Generator" OR "3D model from text" OR "3D Model from image" OR "3D object creation from text" OR "3D object creation from image" in the Scopus database dates from 1996.

The most popular keywords extracted by VOSviewer software (https://www.vosviewer.com/) [13, 14] are three-dimensional computer graphics, 3D modelling, diffusion model, image reconstruction, deep learning, computer vision, 3D reconstruction, image segmentation, semantics, image enhancement, image processing, diffusion, textures, 3D models, generative model, algorithms, rendering, diagnostic imaging, magnetic resonance imaging, artificial intelligence, medical imaging, computed tomography, convolutional neural network, high quality, procedures, learning systems, virtual reality, cameras, nuclear magnetic resonance imaging, geometry, and others. Keywords often appearing in the bibliometric analysis of other topics, such as controlled study, human, male, female, and article, were excluded here. Figure 1 shows a network visualisation of the keywords extracted from 2,621 Scopus documents.

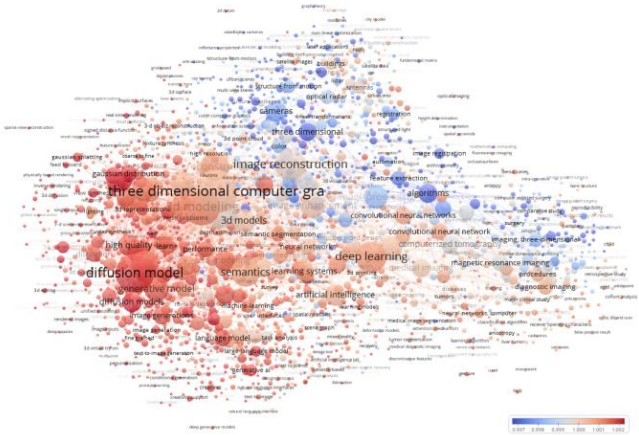

Fig. 1. Network visualisation of keywords extracted from the Scopus database. Colour maps show normalised years from 1996 to 2025. More detailed visualisations can be found at https://youtu.be/FNSRrNTrPxU

## II. ONLINE 3D MODEL GENERATORS

### A. Zoo

The Zoo modelling application can be downloaded from https://zoo.dev/modeling-app/download. Objects can be modelled using the KCL language, which appears to be (or is very similar to) OpenSCAD (https://openscad.org/), a programming language used to create 3D models.

Our two prompts, "rendered Utah teapot" and "Utah teapot with marble texture", produced the same output: a teapot without a lid (Fig. 2). The prompt "Utah teapot cross-section" produced something strange in the form of a plate, apparently a small part of a solid Utah teapot (Fig. 3).

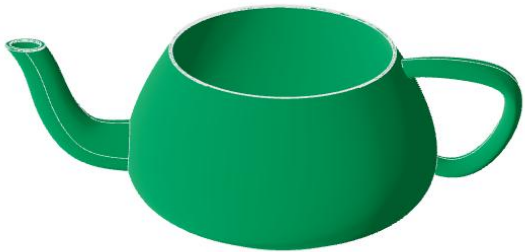

Fig. 2. The Utah teapot.

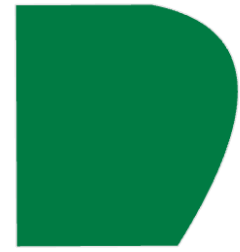

Fig. 3. Cross section of the Utah teapot.

Downloads from the Zoo app are available in gltf, fbx, glb, obj, ply, stl, and step file formats. These files can be opened and processed in CAD or computer graphics software or sent for 3D printing. A brief description of these file types follows.

- FBX: A proprietary file format developed by Autodesk for 3D models, animations and textures, often used in game engines and animation software.
- GLB: A binary version of the GLTF format that transfers 3D models and scenes compactly and efficiently.
- GLTF: A JSON-based format for transmitting 3D models and scenes, optimised for runtime applications and web use.
- OBJ: A widely used, simple text-based format for representing 3D geometry, including vertices, faces and texture mapping.
- PLY: A format commonly used for 3D scanning and point cloud data that can store vertex and face data, colour, and other attributes.
- STEP: A comprehensive CAD file format used for 3D modelling in engineering and product design that stores detailed product data.
- STL: A standard format for 3D printing, representing 3D models using triangular meshes widely supported by 3D printers.

Now, let us critically analyse the aesthetic properties of the model, which is probably stored in the Zoo app's database and has been generated at our request.

The generated geometry's analysis is based on the principles of structural unity of form, proportional consistency, and compositional balance.

In Fig. 2, it can be seen that the generator manages to construct the shape of the main elements of the teapot, consisting of the simplest geometric primitives. It is evident that the constructed volume-space structure of the teapot is generally well organised from the point of view of the connection of the components; the surfaces are smoothed and conjugated according to the $G^0$ criterion since there are no rounded edges at the junctions of the elements. The wall thickness is set correctly enough without any shape distortion. In general, from a technical point of view, the generator fully implements the algorithms of the shape operations that form the basis for constructing the teapot elements.

For example, the teapot's body is constructed using a curved profile followed by the formation of a surface of revolution. The spout and handle are loft objects whose shape is built using a profile and a curved guide.

At the same time, the shape of the teapot should be ergonomic and beautiful. These quality indicators are necessary for the product's content since "most users are interested in the parameters of the device that ensure its convenience, comfort of use, faultless and accurate operation" [15]. In this respect, we note the presence of disproportionate elements in relation to each other and the whole. The spout and handle have the same volume and shape curvature, creating the same visual weight. This makes it challenging to define their functional purpose clearly. In this respect, the overall shape of the teapot feels flattened and horizontally elongated.

Thus, the generated teapot is characterised by sufficiently high-quality surface geometry and non-compliance with consumer quality criteria. An analysis of the teapot's shape allows us to draw a preliminary conclusion that the generator as a whole constructs the shape of the geometric primitives correctly.

*B. NeThing.xyz*

neThing.xyz (https://nething.xyz/) is a coding-based approach to text-to-3D generative AI largely based on CAD. Objects can be created by using programming. Below is an example of a code that generates random objects on a square.

```
# Create a 3D square as the base for other objects
square_base = Box(length=40, width=40, height=2)
# Flat square base
# Add random geometric objects to the square
import random
# Random object creation function
def random_objects_on_square(num_objects=5):
    objects = []
    for _ in range(num_objects):
# Randomise the position on the square base
        x = random.uniform(-20, 20)  # Random x-coordinate within the base width
        y = random.uniform(-20, 20)  # Random y-coordinate within the base length
        z = random.uniform(2, 10)   # Random z-coordinate for elevation off the base
# Randomly select an object type (sphere, cube, or cylinder)
        shape_choice = random.choice(['sphere', 'cube', 'cylinder'])
            if shape_choice == 'sphere':
 # Create a sphere with a random radius
            radius = random.uniform(2, 5)
            sphere = Sphere(radius=radius).translate([x, y, z])
            objects.append(sphere)
              elif shape_choice == 'cube':
            # Create a cube with a random size
            size = random.uniform(3, 6)
            cube = Box(length=size, width=size, height=size).translate([x, y, z])
            objects.append(cube)
        elif shape_choice == 'cylinder':
            # Create a cylinder with a random radius and height
            radius = random.uniform(2, 5)
            height = random.uniform(5, 10)
            cylinder = Cylinder(radius=radius, height=height).translate([x, y, z])
            objects.append(cylinder)
    return objects
# Generate random objects on the square
random_objects = random_objects_on_square(num_objects=10)
# Combine all objects: square base + random objects
scene = square_base + random_objects
# The final result
save = scene
```

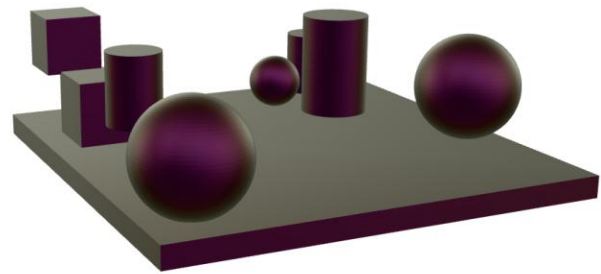

Fig. 4. Random objects on a square.

Figure 4 shows a set of primitives placed on a plane. The generator imposes a particular order on the objects, varies their size and divides them into groups. The objects are correctly differentiated by tone, which allows them to be quickly recognised. An attempt to organise a combination of shapes based on contrasting shapes and sizes creates a certain dynamic and expressiveness. At the same time, from a compositional point of view, the structure is chaotic, with no primary or secondary elements. The colour scheme is limited, based on a combination of disharmonic dirty greens and purples.

*C. Meshy*

Meshy (https://www.meshy.ai/) is an AI 3D model generator that effortlessly turns images and text into 3D models in seconds. It is positioning itself as the number one AI 3D model generator for creators.

The first model we generated using the "Utah teapot" prompt returned an incorrect result, which is not a Utah teapot (Figures 5, 6). This is probably because the model we wanted is missing from the Meshy database. Figure 7 shows the visualisation of the Gaussian curvature on the generated teapot.

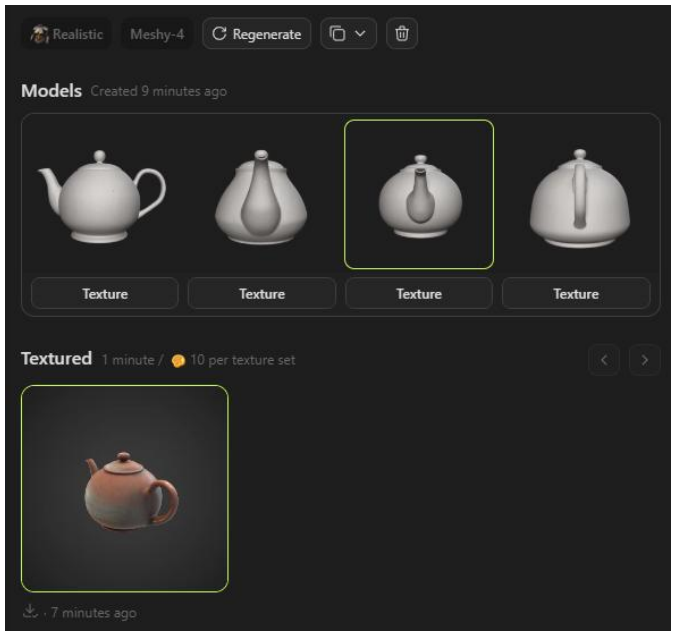

Fig. 5. Four models generated from the text prompt "Utah teapot".

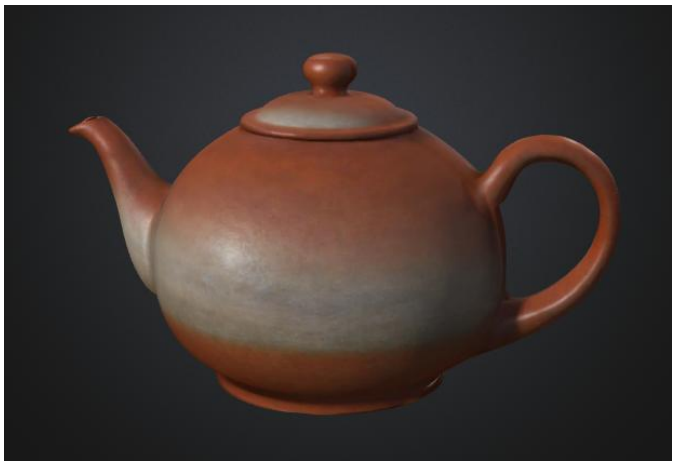

Fig. 6. The selected rendered model.

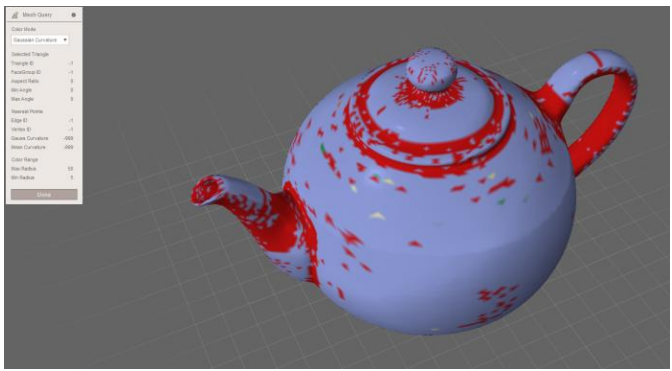

Fig. 7. Visualisation of Gaussian curvature in Meshmixer [16].

Even though the design is far from what we wanted, in the examples with the four generated teapots, it is necessary to note the ability of the generator to achieve a high technical level of design: to perform high-level manipulations with geometry without violating the principles of design and ergonomics. It should be noted that the proportions of the elements in all the pots are close to the proportions of the golden section. The components are consistent in size, shape and style. Their position meets the requirements of comfort and safety.

All these features make it easier to understand the functional role of each teapot component and help create a memorable image of the product. In the colour model of the teapots, the generator reproduces the colour of the clay with simulated burnout. It is important to note that the positions of the bright areas of the clay texture are not random but dictated by the need to identify the shape features of the components and highlight their functional boundaries.

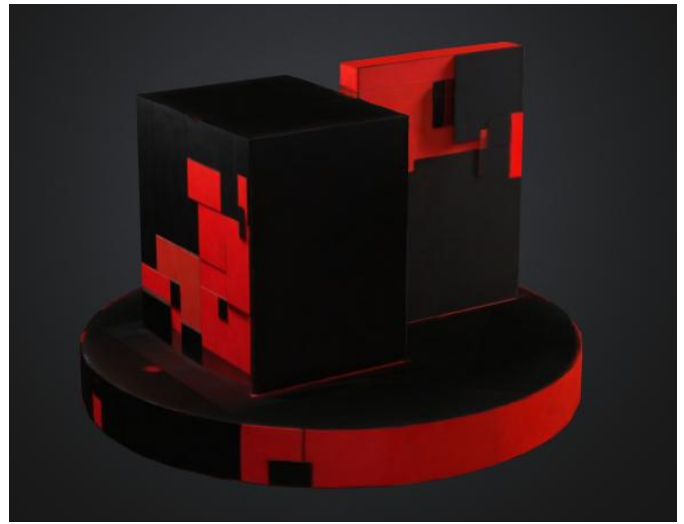

Fig. 8. 3D black and 3D red squares based on corresponding paintings by Kasemir Malevich.

Figure 8 presents a 3D interpretation of the famous paintings 'Black Square' and 'Red Square' by Kazemir Malevich [17,18], whose pioneering work and writings influenced the development of abstract art in the 20th century. In the three-dimensional structure, the relationships between the elements in terms of shape, size and contrast are expressed quite competently. There is a supporting base in the form of a cylindrical disc. It is opposed in shape by two supporting parallelepipeds. Their shape is the same, but the contrast in size makes it possible to determine their role in the overall composition: the black parallelepiped is the centre of the composition, and the smaller one is subordinate. In this way, the generator skilfully creates a composite structure.

An attempt is made to apply a red texture to the surface, combining the size of the graphics and thereby playing out the shape of the elements. It should also be noted that the generator uses colour patches of different sizes to create a visual balance and reveal the form's geometry. In the formal structure, however, there is an inconsistency in the size of the cylindrical disc in relation to the supported elements: its thickness could be less since, in terms of visual mass, it competes with the visual mass of two parallelepipeds. With a smaller thickness, the visual activity of the cylindrical disc will decrease, and the activity of the supported elements will increase. Generally, the 3D interpretation of the artist's paintings is characterised by expressiveness and competent composition.

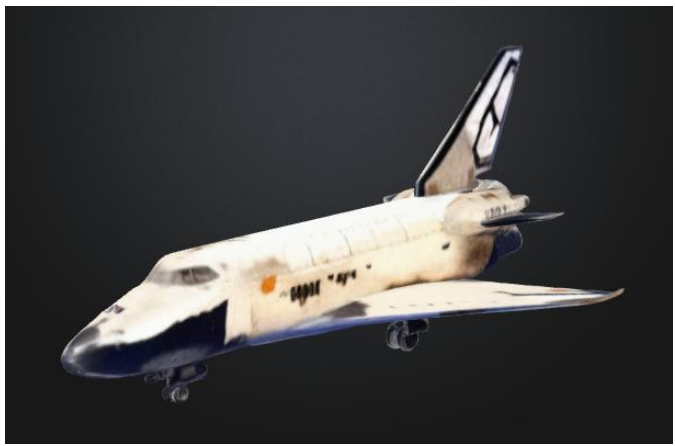

Fig. 9. A Soviet "Buran" space shuttle.

Figure 9 shows an image of the Soviet space shuttle "Buran" [19] generated from the provided image. The overall shape correctly describes the image of the original object but is characterised by poor surface geometry. In this respect, the object lacks the detail typical of low-poly objects. Similarly, the program uses a low-resolution texture from a used image to simulate the burnt hull of a shuttle.

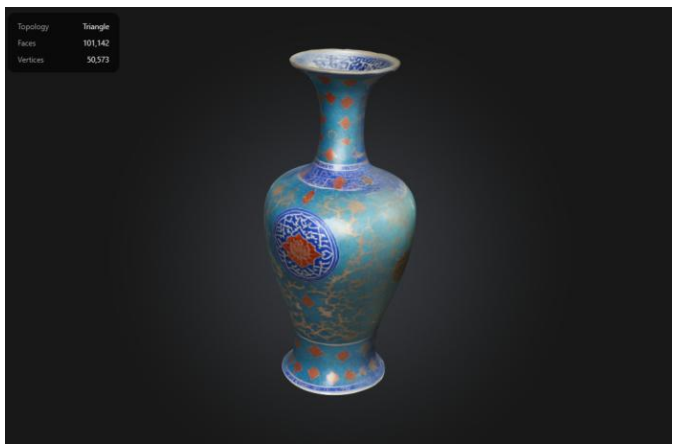

Fig. 10. A Chinese vase.

As mentioned above, the software is good at generating surfaces of revolution, and the Chinese vase (Fig. 10) is no exception in this respect. The shape of the vase meets the criteria of proportional consistency, as well as the traditional Chinese canons of shape and stylisation. The texture pattern is correctly applied to the surface, and its elements are legible. The drawing pattern meets compositional balance requirements, as the primary and secondary elements are clearly readable. The colour composition is based on the rules of colour harmony, a contrasting and nuanced combination of shades, using a group of warm and cold colours.

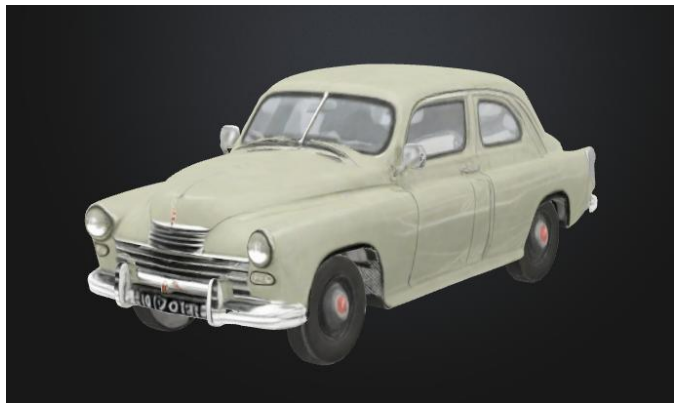

Fig. 11. A Soviet car model "Pobeda".

The Soviet "Pobeda" car [20] model, created in 1946, is also a low-poly object with more detail. The elements of the car's shape are easier to read. The program simulates various materials and their physical properties through reflections and glare. The model's colour scheme is based on a combination of a low-saturation light green body colour with a dark grey wheel colour. There are no apparent errors in the layout of the car elements.

### III. CONCLUSIONS

This work aimed to investigate various online tools for text-to-3D and image-to-3D generators and analyse the generated models critically. The results of this critical case study may be helpful for software developers focusing on AI-based 3D model generation and related fields.